\documentclass[aps,prb,twocolumn,superscriptaddress,showpacs,floatfix,10pt]{revtex4-2}

\usepackage{amsmath}
\usepackage{amsfonts}
\usepackage{graphicx}
\usepackage{color}
\usepackage{epstopdf}
\usepackage{natbib}
\usepackage{xfrac}
\usepackage{enumitem}
\usepackage[english]{babel}
\usepackage{booktabs}
\usepackage{soul}
\usepackage{xcolor}
\usepackage{lipsum}
\usepackage{tabularx}
\usepackage{multirow} 

\usepackage{float} 
\usepackage{subfigure}

\usepackage{booktabs}
\usepackage{multirow}
\usepackage{comment}

\begin{document}
\title{
Complex Refractive Index Extraction for Spintronic Terahertz Emitter Analysis
}

\author{Yingshu Yang}
\email{yingshu.yang@ntu.edu.sg}
\affiliation{Division of Physics and Applied Physics, School of Physical and Mathematical Sciences (SPMS), Nanyang Technological University, 21 Nanyang Link, Singapore 637371, Singapore}

\author{Keynesh Dongol}
\affiliation{Department of Materials Science and Engineering, University of Wisconsin-Madison, Madison, WI, 53706, USA}

\author{Stefano {Dal~Forno}}
\affiliation{Division of Physics and Applied Physics, School of Physical and Mathematical Sciences (SPMS), Nanyang Technological University, 21 Nanyang Link, Singapore 637371, Singapore}

\author{Ziqi Li}
\affiliation{Division of Physics and Applied Physics, School of Physical and Mathematical Sciences (SPMS), Nanyang Technological University, 21 Nanyang Link, Singapore 637371, Singapore}

\author{Piyush Agarwal}
\affiliation{Institute of Materials Research and Engineering (IMRE),
Agency for Science Technology and Research(A*STAR), 2 Fusionopolis Way, Innovis 08-03, Singapore138634, Singapore}

\author{Amalini Mansor}
\affiliation{Division of Physics and Applied Physics, School of Physical and Mathematical Sciences (SPMS), Nanyang Technological University, 21 Nanyang Link, Singapore 637371, Singapore}

\author{Ranjan Singh}
\affiliation{Department of Electrical Engineering, University of Notre Dame, Notre Dame, IN, 46556 USA}

\author{Marco Battiato}
\email{marco.battiato@ntu.edu.sg}
\affiliation{Division of Physics and Applied Physics, School of Physical and Mathematical Sciences (SPMS), Nanyang Technological University, 21 Nanyang Link, Singapore 637371, Singapore}

\author{Elbert E. M. Chia}
\email{ElbertChia@ntu.edu.sg}
\affiliation{Division of Physics and Applied Physics, School of Physical and Mathematical Sciences (SPMS), Nanyang Technological University, 21 Nanyang Link, Singapore 637371, Singapore}

\author{Guoqing Chang}
\email{guoqing.chang@ntu.edu.sg}
\affiliation{Division of Physics and Applied Physics, School of Physical and Mathematical Sciences (SPMS), Nanyang Technological University, 21 Nanyang Link, Singapore 637371, Singapore}

\date{\today}

\begin{abstract}

Spintronic terahertz emitters (STEs) generate broadband terahertz (THz) radiation, which is essential for spectroscopy, imaging, and communication.
The performances and the essential physical parameters of STE devices are linked to the dielectric properties of the constituent materials.
Terahertz time-domain spectroscopy (THz-TDS) is an effective tool to measure these properties, but conventional analysis struggles with thin or complex multilayered systems due to simplifying approximations or complex transfer functions. In this work, we present a practical method to extract dielectric properties of STE multilayers using the Transfer Matrix Method (TMM).
By comparing the THz pulse calculated using the Transfer Matrix Method (TMM) with the experimentally measured pulse transmitted through the sample, we can extract the dielectric properties of STEs, enhancing THz-TDS analysis and facilitating STE design and optimization.
This method avoids constructing complex transfer functions, accommodates diverse sample geometries, and is designed to be accessible, with a publicly available codebase, making it a useful tool for STE research.

\end{abstract}


\maketitle

\section{Introduction}

Spintronic Terahertz emitters (STEs) are cutting-edge devices that use the interaction between spin currents and magnetization dynamics to produce broadband terahertz (THz) radiation, covering a frequency range from 0.1 to 30 THz \cite{seifert2022spintronic,yang2024theoretical}. This range is crucial for applications in imaging \cite{chen2020ghost,luo2020nanoscale}, spectroscopy \cite{beard2002terahertz}, and wireless communication \cite{koenig2013wireless}. The underlying principle of STEs relies on the conversion of spin current into a transient charge current, facilitated by mechanisms such as the inverse spin Hall effect (ISHE) or the inverse Rashba-Edelstein effect (IREE). These effects drive the generation of THz radiation when a femtosecond laser pulse excites a ferromagnet/non-magnet heterostructure, causing spin currents to flow from the ferromagnetic layer to the non-magnetic layer, where ultrafast charge dynamics generate the THz pulses \cite{kampfrath2013terahertz}.

The rise of STEs has gained attention due to their ability to efficiently produce broadband, tunable THz pulses \cite{seifert2022spintronic,cheng2019far}. Compared to traditional THz emitters like photoconductive antennas or optical rectifiers, STEs offer significant advantages in terms of bandwidth, efficiency, and scalability \cite{bull2021spintronic}. Their compact nature allows for easier integration into existing electronic and spintronic systems, making them a flexible tool in advancing THz technology.

\begin{figure}[!t]
\centering
\includegraphics[width=\linewidth]{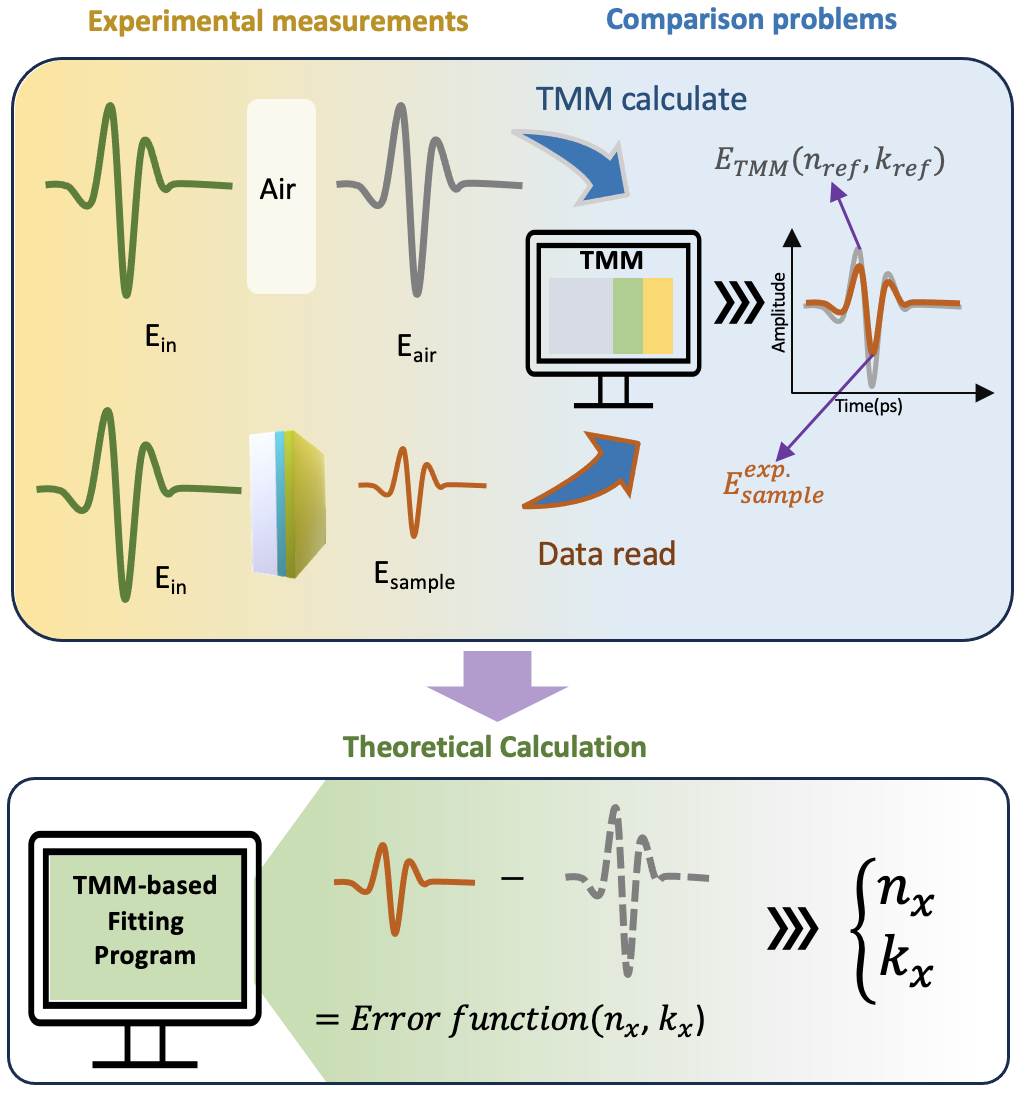}
\caption{The description of the variation problem when directly using the database for calculation and the recipe of the TMM-based dielectric property extraction. }
\label{fig:fig1}
\end{figure}

Recently, the research in this field is focusing on optimizing STE performance by enhancing the efficiency, bandwidth, and tunability of THz emission \cite{yang2024theoretical,seifert2022spintronic}. Significant progress has come from exploring material systems like magnetic multilayers and heterostructures \cite{seifert2022spintronic}. In these cases, researchers have used the Transfer Matrix Method with Source (TMMS) to model charge current and THz emission processes \cite{yang2021transfer,agarwal2023secondary}. General models for thickness-dependent THz emission amplitudes have also been applied \cite{seifert2016efficient,qiu2018layer,torosyan2018optimized,zhou2018broadband,yang2024theoretical}. However, the precision of these models relies on the accuracy used to describe the dielectric properties of the materials\cite{yang2023modeling}. While standard values of the dielectric constants exist in databases, experimental conditions often lead to notable variations (shown in Fig.~\ref{fig:fig1} comparison problems).
Moreover, several other physical parameters such as spin Hall angle, diffusion lengths and the like, are often difficult to measure \cite{yang2023modeling,liu2022spintronic,agarwal2023secondary,cheng2019far}.
For this reasons, developing efficient methods to extract these values from experiments is essential for a reliable modeling and optimization of STE devices \cite{yang2024theoretical}.

To accurately characterize the material properties of STEs (typically thin multilayered samples), THz time-domain spectroscopy (THz-TDS) has proven to be a powerful technique. THz-TDS enables a detailed study of how THz pulses interact with materials, though analyzing the resulting data presents challenges \cite{smith2011terahertz,neu2018tutorial}. A key aspect of this analysis is the transfer function, defined as the complex ratio of the sample’s Fourier-transformed complex THz electric field ($\tilde{E}_{sample}$) to that of one measured with a reference sample ($\tilde{E}_{reference}$) \cite{tayvah2021nelly,wilk2008highly,duvillaret1996reliable,duvillaret1999highly,pupeza2007highly,beermann2025terahertz}. This function captures essential information about how the material modifies the THz pulse as it passes through.

For simple, thick-layered samples, calculating the transfer function is straightforward, often yielding closed-form expressions that simplify data interpretation. However, thin multilayered samples, common in STEs, introduce significant complexity, necessitating advanced analysis methods. In such cases, the transfer function’s complexity increases dramatically \cite{duvillaret1996reliable,tayvah2021nelly}.
Traditional approaches often rely on approximations, such as neglecting certain reflections or using the Tinkham formula via Taylor expansion\citep{tayvah2021nelly}. These simplifications, while useful, depend on specific assumptions about sample properties, limiting their applicability and potentially leading to unreliable results if misapplied \citep{tayvah2021nelly}. Alternative methods, such as tree structures map all possible THz pulse paths and construct the transfer function without simplifying assumptions. This avoids approximations, but requires sophisticated and computationally intensive algorithms \citep{cassar2019iterative,tayvah2021nelly}.
Additionally, for optically thick samples, echo-resolved THz spectroscopy has been employed to simultaneously determine complex optical constants and sample thickness by numerically solving transcendental transmission equations from main and echo pulses \cite{beermann2025terahertz}.  While effective for thicker samples, these methods are less suited for the thin, multilayered systems typical of STEs, where thin multilayers and substrate contributions complicate the analysis.

The Transfer Matrix Method (TMM) offers a robust framework for modeling electromagnetic (EM) wave transmission through multilayered samples, capturing all Fabry-Perot reflections in a single matrix \cite{wilk2008highly}. However, THz-TDS measurements typically use a finite time window, while TMM assumes an infinite one \cite{wilk2008highly}.
For thin layers, where reflections overlap quickly, a finite window will work. However, in STEs, the thin layers often sit on thick substrates. The substrate will add delayed reflections that demand a longer measurement requiring the infinite window, which the TMM expects. This mismatch has long blocked the use of TMM to extract THz material properties \cite{yang2021removal,wilk2008highly}. 

To address this, we present a practical TMM-based method to extract dielectric properties of thin multilayer samples, as outlined in Fig.~\ref{fig:fig1}. By incorporating an analytical modification to account for thick substrates \citep{yang2021removal}, our approach uses the measured THz electric field through dry air ($E_{air}(t)$) as a reference, eliminating the need for an additional reference sample. This method avoids complex tree structures \citep{cassar2019iterative,tayvah2021nelly} and is designed to be lightweight and accessible, with a publicly available codebase and a simple graphical user interface (GUI).

Our approach is not intended to replace specialized material parameter fitting tools \citep{duvillaret1996reliable,duvillaret1999highly,tayvah2021nelly}, but rather to provide a practical tool tailored for STE analysis. It bridges experimental THz-TDS data with STE modeling, enabling researchers to extract dielectric properties directly from raw time-domain data for use in TMM-based models like Transfer Matrix Method with Source (TMMS)\citep{yang2021transfer} or the Perturbative Transfer Matrix Method (PTMM) \citep{yang2024perturbative}.

\begin{figure}[!t]
\centering
\includegraphics[width=\linewidth]{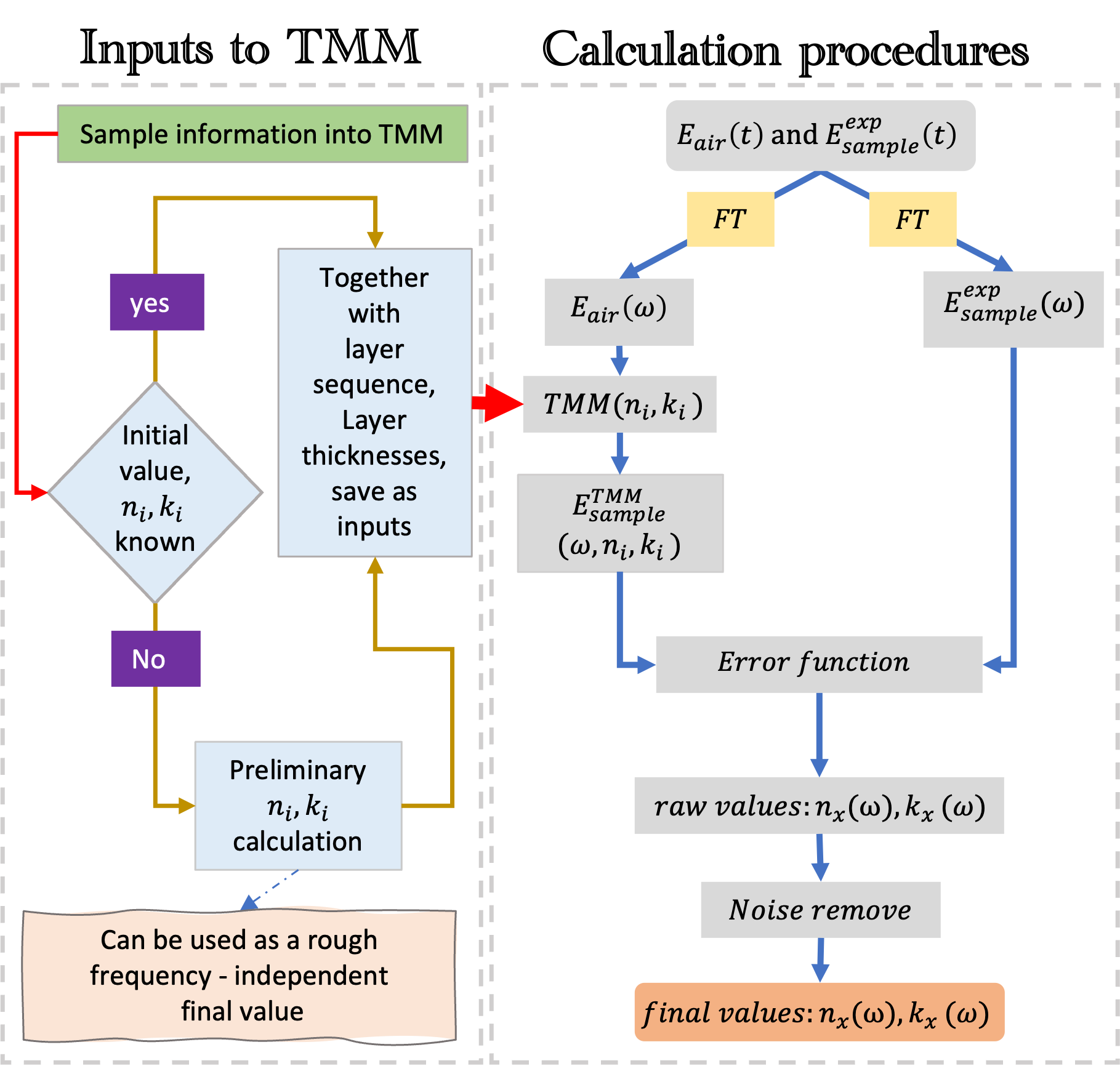}
\caption{Block diagram illustrating the key stages of the TMM-based dielectric property extraction process.}
\label{fig:fig2}
\end{figure}

\begin{figure}[ht]
\centering
\includegraphics[width=1.0\linewidth]{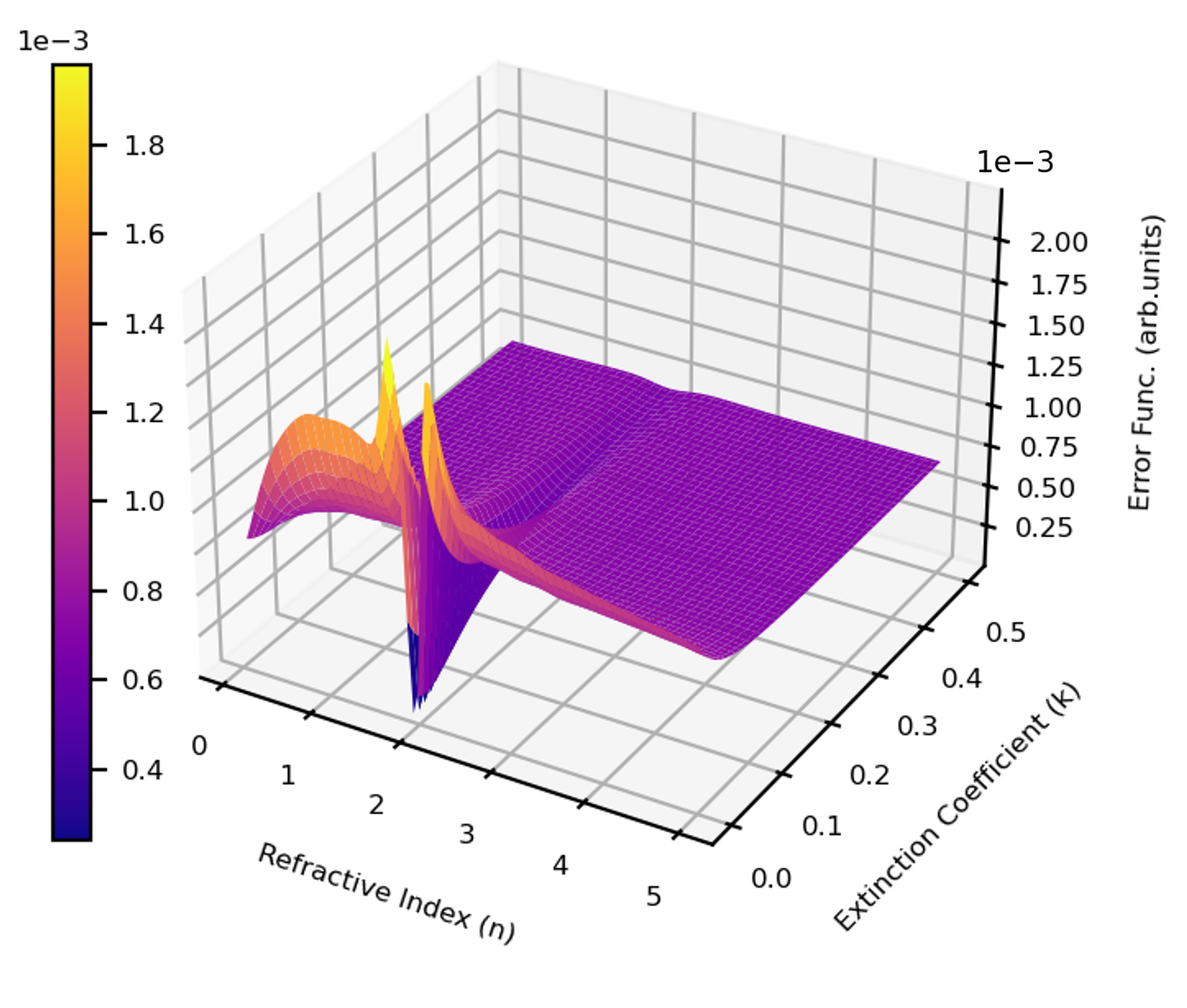}
\caption{Surface plot of the error function used in TMM fitting for dielectric constant extraction. The figure shows the error landscape as a function of refractive index ($n$) and extinction coefficient ($k$) for 1mm thick quartz. Color indicates the magnitude of the error in arbitrary units. The error minima identify the optimal $(n, k)$ pairs for initializing frequency-dependent fitting and for estimating material parameters.
}
\label{fig:plot3}
\end{figure}

\section{Methods}

\subsection{Overview of the Modified Transfer Matrix Method}

This section briefly reviews the TMM and the echo removal technique \cite{yang2021removal}, which are essential for analyzing EM wave propagation through layered media. The derivation, using notation consistent with this work, is detailed in Ref.~\cite{yang2021removal}. The TMM provides a solid way to tackle EM waves in layered systems. By applying continuity conditions at interfaces, the system’s behavior can be captured in a matrix form,
\begin{equation}\label{eq:T_nm}
    \mathbf{T}_{[n,m]}=\mathbf{A}_{m}^{-1}\left[0 \right] \left(\prod_{j=m-1}^{n+1} \mathbf{A}_j[d_j] \mathbf{A}_j^{-1}[0] \right)\mathbf{A}_n\left[d_n \right], 
\end{equation}
where $\mathbf{T}_{[n,m]}$ is the transfer matrix between layers $n$ and $m$ (with $n < m$), $d_n$ denotes the thickness of the $n$-th layer, and $\mathbf{A}_n$ represents the layer's characteristic matrix:
\begin{align}
    &\mathbf{A}_n\left[\omega,z\right]=\begin{bmatrix} e^{i \omega \sqrt{\epsilon_n \mu_n} z } &e^{-i \omega \sqrt{\epsilon_n \mu_n} z} \\ 
    -\sqrt{\frac{\epsilon_n}{\mu_n}}e^{i \omega \sqrt{\epsilon_n \mu_n} z }&{\sqrt{\frac{\epsilon_n}{\mu_n}} e^{-i \omega \sqrt{\epsilon_n \mu_n} z}}\end{bmatrix},
\end{align}
with $\epsilon_n$ and $\mu_n$ being the permittivity and permeability of the $n$-th layer, respectively, and $z$ the position within the layer.

To model transmission experiments, we ensure no left-propagating wave exists in the terminal air layer ($N$), yielding:
\begin{equation}
	\begin{bmatrix} f_{N}^>\\ 0\end{bmatrix} = \mathbf{T}_{[0,N]} \begin{bmatrix} f_{0}^>\\  f_{0}^<\end{bmatrix}.
\end{equation}
Here, $f_{0}^>$ and $f_{0}^<$ signify the incident and reflected waves at the initial interface, while $f_{N}^>$ represents the wave transmitted through to the last layer. The coefficients of transmission ($t$) and reflection ($r$) for the entire system are defined as:
\begin{align}
    t_{[0,N]} &=\frac{f_{N}^>}{f_{0}^>}, \\
    r_{[0,N]}&=\frac{f_{0}^<}{f_{0}^>},
\end{align}
derived from the system's total transfer matrix $\mathbf{T}_{[0,N]}$. 
To remove the substrate influence, TMM calculations are split into two steps: first across the air/substrate interface, then through the subsequent layers, and vice versa. This approach allows the transmission through the entire multilayer system to be expressed as:
\begin{align}
    f_{N}^>&=t_{[0,S]} t_{[S,N]}\; f_{0}^>,
\end{align}
where $S$ denotes the substrate layer index.

\begin{figure*}[ht]
    \centering
    \includegraphics[width=1\textwidth]{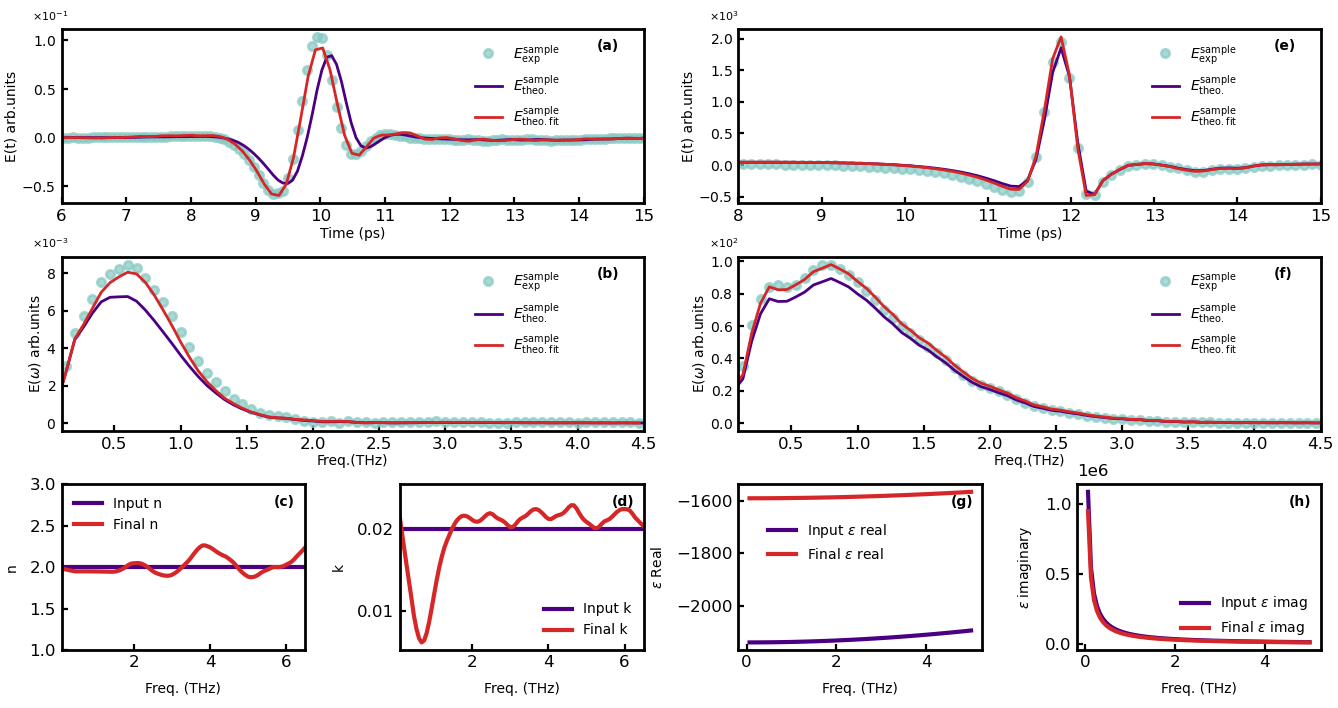}
    \caption{
    Terahertz (THz) propagation analysis for two samples: quartz (1\,mm) and Pt (6\,nm)/sapphire (500\,$\mu$m). Panels (a–b) show results for the quartz sample, while panels (e–f) correspond to the Pt/sapphire heterostructure. Time-domain electric field waveforms are shown in panels (a) and (e), and their respective frequency-domain spectra are shown in panels (b) and (f). Experimental data are represented by cyan dots ($E_{\text{exp}}^{\text{sample}}$), theoretical calculations using literature dielectric constants from Ref.~\cite{franta2016optical} (quartz) and Ref.~\cite{ordal1985optical} (Pt) are shown as purple curves ($E_{\text{theo}}^{\text{sample}}$), and the fitted results obtained via the TMM are shown as red curves ($E_{\text{theo,fit}}^{\text{sample}}$). Panels (c–d) present the literature input (purple) and TMM-extracted (red) optical constants $n$ and $k$ for quartz. Panels (g–h) show the literature input and TMM-fitted real and imaginary parts of the permittivity $\varepsilon$ for the Pt layer (calculated using Drude model). These results demonstrate the refinement of optical parameters through TMM fitting to accurately reproduce experimental THz transmission.
    }
    \label{fig:plot}
\end{figure*}

\subsection{Implementation}

\textbf{Sample Geometry.} STEs consist of a series of layers with specific thicknesses, typically arranged as air, a substrate, a first layer, a second layer, and air again. Each layer has a refractive index $n$ and extinction coefficient $k$, which describe its dielectric properties. For the substrate, we know these values ($n_{sub}$ and $k_{sub}$), as well as the thickness of each layer. However, for one or more layers, $n$ and $k$ are unknown, these are the free parameters we aim to find using our method.

\textbf{Measurements and Calculations.} We find these unknown dielectric properties through simple experimental measurements and numerical calculations, outlined in the following steps:

1. \textit{Baseline Measurement:} We measure the THz pulse passing through dry air (no sample present) in the time domain  and record it as $E_{air}(t)$. 

2. \textit{Sample Measurement:} We measure the THz pulse after it passes through the sample in time domain, labeling this as $E_{sample}^{exp}(t)$.

3. \textit{Theoretical Simulation:} We use the TMM to calculate the theoretical transmission through the sample, called $E_{sample}^{TMM} (t)$, using $E_{air} (t)$ as the input. (Note that a time delay often appears between the measured and calculated THz pulses due to the pulse’s travel time through air. We account for and correct this delay in our calculations.)
The TMM needs the thickness and dielectric properties of each layer. Since the unknown layer’s $n$ and $k$ aren’t known, we start with guessed values.

4. \textit{Numerical Fitting:} Our method estimates the unknown $n$ and $k$ by comparing the TMM’s predicted transmission to the experimental measurement. We use the Nelder-Mead algorithm to minimize the difference between $\tilde{E}_{sample}^{TMM} (\omega)$ and $\tilde{E}_{sample}^{exp}(\omega)$. The total error function used in the optimization routine is defined as the sum of squared differences between the experimentally measured THz signal in the frequency domain and the theoretical signal computed via the TMM:
\begin{equation}
\epsilon_{err} = \sum_{\omega_i} \left| \tilde{E}_{\text{sample}}^{\text{exp}}(\omega_i) - \tilde{E}_{\text{sample}}^{\text{TMM}}(\omega_i; n(\omega_i), k(\omega_i)) \right|^2.
\end{equation}
Here, \( \tilde{E}_{\text{sample}}^{\text{exp}} \) is the complex electric field in the frequency domain, obtained by Fourier transforming the experimentally measured time-domain signal. \( \tilde{E}_{\text{sample}}^{\text{TMM}} \) is the simulated transmitted field computed using the TMM. The TMM calculation is performed independently at each frequency point, using either a complex permittivity derived from the Drude model (for metallic layers such as Pt or Fe), or from frequency-dependent refractive index $n(\omega)$ and extinction coefficient $k(\omega)$values, depending on the parameterization selected by the user.

Fig.~\ref{fig:fig2} illustrates a block diagram outlining the key stages of our dielectric constant extraction method using the TMM. The procedure begins by inputting sample-specific information, including layer sequence and thicknesses, followed by the selection or estimation of initial optical constants ($n$, $k$). If initial values are unknown, a preliminary estimation can be obtained via a rapid scan of the error function landscape, which provides a frequency-independent approximation suitable for applications not requiring spectral resolution. Further algorithmic details and input specifications are provided in the Supplementary Material \cite{suppinfo}, which also include Refs. \cite{seifert2018terahertz, ordal1983optical,blaber2009search,zeman1987accurate,hooper2002dispersion,grady2004influence}.

As a demonstration of the preliminary estimation step, Fig.~\ref{fig:plot3} presents a surface plot of the TMM error function for a representative sample, 1\,mm quartz. The plot illustrates the error magnitude as a function of the refractive index ($n$) and extinction coefficient ($k$), with the global minimum indicating the optimal $(n, k)$ pair. In this case, we have applied a frequency-independent error function:
\begin{equation}
\epsilon_{err}(n, k) = \sum_{\omega} \left| \tilde{E}^{\text{exp}}_{\text{sample}}(\omega) - \tilde{E}^{\text{TMM}}_{\text{sample}}(\omega; n, k) \right|^2.
\end{equation}
This method provides a fast and effective mean for coarse fitting when frequency-independent values are sufficient, and also serves as a robust initialization strategy for more detailed, frequency-dependent optimization routines.


\begin{figure}[htbp]
    \centering
    \includegraphics[width=\linewidth]{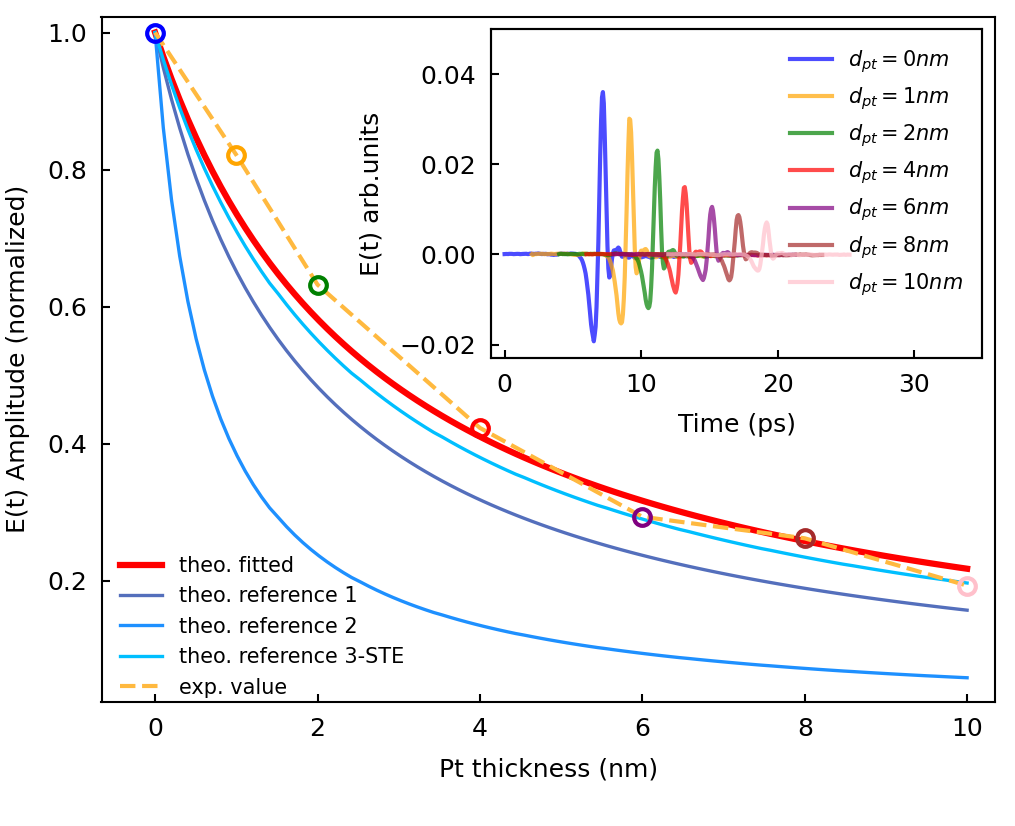}
    \caption{THz transmission through NiFe(3nm)/Pt(x-nm)/Quartz(1mm) as a function of Pt thickness. 
    \textbf{Main panel:} normalized peak-to-peak amplitude of the transmitted THz field versus \(d_{\mathrm{Pt}}\) (0–10\,nm). Circles are measurements; the dashed line is a guide to the eye. Solid curves are transfer-matrix simulations using (i) Pt dielectric parameters from the literatures \cite{ordal1985optical,rakic1998optical, seifert2018terahertz} (blue lines), and (ii) Pt Drude parameters obtained from our fit to the transmission data (red). For comparison, a simulation using Drude parameters extracted independently from spintronic-emitter (STE) samples (“reference-3-STE”, light blue) is also shown. 
    \textbf{Inset:} representative time-domain THz waveforms for several \(d_{\mathrm{Pt}}\); traces are horizontally offset for clarity and illustrate the progressive attenuation with increasing Pt thickness.}
    \label{fig:plot2}
\end{figure}

\section{Results discussion}
Our method is useful for modeling the dielectric constant across a wide range of materials and conditions. However, we focus primarily on structures with multiple thin layers on a thick substrate. A key application of the dielectric constant at THz frequencies in these systems is analyzing the outcoupling of THz emissions in STEs. In STEs, THz emission is generated internally and then transmitted through the material, often serving as a probe for understanding THz generation mechanisms. This highlights the importance of precise material property.

Our data extraction is thus tailored to geometries featuring a thick substrate topped with thin layers, a configuration where simplified methods often struggle due to the difficulty of deriving a closed-form transfer function. 

\subsection{Impact on single thickness samples}

To evaluate the effectiveness of our approach in characterizing STEs, we compare experimentally measured THz transmission signals with numerically simulated results based on both literature and extracted dielectric constants, as shown in Fig.~\ref{fig:plot}. For quartz (Figs.~\ref{fig:plot}a–b) and Pt/sapphire samples (Figs.~\ref{fig:plot}e–f), the initial simulations (purple curves), which rely on dielectric constants from literature sources~\cite{franta2016optical,ordal1985optical}, show noticeable discrepancies with the experimental data (cyan dots). This deviation highlights the limitations of using generic dielectric constants, which may not accurately capture the true response of materials as fabricated and measured in specific experimental setups.

In contrast, simulations based on dielectric constants extracted using our TMM approach (red curves) exhibit significantly improved agreement with the experimental measurements. For the quartz sample, the extracted refractive index ($n$) and extinction coefficient ($k$) (Figs.~\ref{fig:plot}c–d) differ notably from the literature values, indicating the importance of sample-specific characterization. For the Pt/sapphire heterostructure, the real and imaginary parts of the permittivity $\varepsilon$ (Figs.~\ref{fig:plot}g–h) were calculated using a Drude model with fitted plasma and damping frequencies, yielding better alignment with the observed THz response (for a detailed description of the Drude model used, please refer to Section 5 of the Supplementary information \cite{suppinfo}). These results underscore the limitations of relying solely on tabulated dielectric constants and highlight the robustness of our method in extracting accurate, material-specific optical parameters critical for reliable THz pulse simulation and modeling in STE systems. For additional testing results and a detailed description of the instructions in using the TMM program, please refer to the Supplementary Materials\cite{suppinfo}.

\subsection{Impact on changing thickness samples}

In addition to single-thickness validation, we extended our analysis to a series of samples with varying Pt layer thicknesses, a common approach in the study of STEs. Such thickness-dependent measurements are frequently used to extract key performance indicators, including the spin Hall angle and spin diffusion length of the layer materials used ~\cite{seifert2016efficient,yang2024theoretical,liu2022spintronic,agarwal2023secondary}.

Fig.~\ref{fig:plot2} presents the normalized transmitted THz field amplitude as a function of Pt thickness for NiFe(3nm)/Pt(x-nm)/Quartz(1mm) samples. The experimental points (circles; dashed guide-to-the-eye) decay with increasing thickness of Pt layer. Simulations based on literature dielectric parameters for Pt\cite{ordal1985optical,rakic1998optical}, underestimate transmission amplitudes over the full thickness range, indicating that standard optical constants over-penalize losses in our samples and measurement conditions. By contrast, the curve obtained with Pt Drude parameters extracted from our TMM fit to the transmission data (red) closely follows the experimental trend across all thicknesses. 
Notably, the “reference-3-STE” curve (light blue) uses Drude parameters obtained from STE samples and already captures much of the observed behavior, supporting the consistency and transferability of our extracted parameters across distinct experimental modalities. The inset displays time-domain THz traces for representative Pt thickness, the progressive attenuation with thickness mirrors the amplitude decay in the main panel.

The optimized Pt parameters inferred here provide a more faithful basis for modeling thickness-dependent THz emission experiments. In particular, when combined with standard emission formalisms, they enable reliable determination of the optimal Pt thickness and more accurate extraction of key transport quantities (e.g., spin-diffusion length) from thickness-dependent THz signals.

While numerous methods exist for extracting optical parameters from THz transmission data\cite{dorney2001material,wilk2008highly,duvillaret1999highly,tayvah2021nelly,la2017phonon}, a direct benchmarking against those techniques is beyond the scope of this study. This work aims to demonstrate the validity and effectiveness of our TMM approach through close agreement with experimental data across various sample configurations in STE studies. We thus focus here on establishing the internal consistency and practical applicability of our method, with comparative studies reserved for future work aimed at broader methodological evaluation. 
{Furthermore, we remind that in applying the modified TMM, we assume ideal interfaces between layers to streamline the extraction of core dielectric properties, a simplification validated by the high-quality STE samples characterized in the Supplementary Information \cite{suppinfo}. Heterostructures with rougher interfaces may introduce scattering losses, potentially leading to systematic errors in the fitted refractive indices and extinction coefficients. For samples with rougher interfaces, future refinements could incorporate interface characterization data (e.g., from AFM or XRR) into more advanced models to enhance precision. Here, we focus on demonstrating the effectiveness of the core TMM-based method.

\section{conclusion}

In this work, we tackled key challenges in characterizing the dielectric properties of multilayered structures in STEs by developing an approximation-free method based on the TMM. Unlike traditional approaches, our method avoids simplifications and handles arbitrary sample geometries, enhancing the accuracy of THz-TDS data analysis. By simulating single- and varying-thickness layers, we showed how our approach effectively extracts material properties essential for optimizing STE performance.
Our findings emphasize the need for precise material property extraction to simulate and understand STE behavior. Using just an initial input pulse and a transmitted pulse measurement, our method offers a solid framework for studying thin or complex multilayer systems, making it especially useful for advancing spintronic technologies.
This work strengthens THz-TDS as a tool for probing spintronic materials and paves the way for future improvements in STE design and applications.
 
\section{Acknowledgment}

This work was supported by the Singapore Ministry of Education Academic Research Fund (AcRF) Tier 2 Grant “Dynamical Characterization of Spin-to-Charge Conversion Mechanisms in 2D Heterostructures” (MOE-T2EP50222-0014), and the Tier 3 Grant “Quantum Geometric Advantage” (MOE-MOET32023-0003).

The data that support the findings of this article are openly available \cite{agarwal2022terahertz,li2025electric}.

\bibliography{apssamp}

@article{smith2011terahertz,
  title={Terahertz time-domain spectroscopy of solid samples: principles, applications, and challenges},
  author={Smith, Ryan M and Arnold, Mark A},
  journal={Applied Spectroscopy Reviews},
  volume={46},
  number={8},
  pages={636--679},
  year={2011},
  publisher={Taylor \& Francis}
}

@article{neu2018tutorial,
  title={Tutorial: An introduction to terahertz time domain spectroscopy (THz-TDS)},
  author={Neu, Jens and Schmuttenmaer, Charles A},
  journal={Journal of Applied Physics},
  volume={124},
  number={23},
  year={2018},
  publisher={AIP Publishing}
}

@article{seifert2022spintronic,
  title={Spintronic sources of ultrashort terahertz electromagnetic pulses},
  author={Seifert, Tom S and Cheng, Liang and Wei, Zhengxing and Kampfrath, Tobias and Qi, Jingbo},
  journal={Applied Physics Letters},
  volume={120},
  number={18},
  year={2022},
  publisher={AIP Publishing}
}

@article{tayvah2021nelly,
  title={Nelly: A user-friendly and open-source implementation of tree-based complex refractive index analysis for terahertz spectroscopy},
  author={Tayvah, Uriel and Spies, Jacob A and Neu, Jens and Schmuttenmaer, Charles A},
  journal={Analytical Chemistry},
  volume={93},
  number={32},
  pages={11243--11250},
  year={2021},
  publisher={ACS Publications}
}

@article{cassar2019iterative,
  title={Iterative tree algorithm to evaluate terahertz signal contribution of specific optical paths within multilayered materials},
  author={Cassar, Quentin and Chopard, Adrien and Fauquet, Frederic and Guillet, Jean-Paul and Pan, Mingming and Perraud, Jean-Baptiste and Mounaix, Patrick},
  journal={IEEE Transactions on Terahertz Science and Technology},
  volume={9},
  number={6},
  pages={684--694},
  year={2019},
  publisher={IEEE}
}

@article{duvillaret1996reliable,
  title={A reliable method for extraction of material parameters in terahertz time-domain spectroscopy},
  author={Duvillaret, Lionel and Garet, Frederic and Coutaz, J-L},
  journal={IEEE Journal of selected topics in quantum electronics},
  volume={2},
  number={3},
  pages={739--746},
  year={1996},
  publisher={IEEE}
}

@article{duvillaret1999highly,
  title={Highly precise determination of optical constants and sample thickness in terahertz time-domain spectroscopy},
  author={Duvillaret, Lionel and Garet, Fr{\'e}d{\'e}ric and Coutaz, Jean-Louis},
  journal={Applied optics},
  volume={38},
  number={2},
  pages={409--415},
  year={1999},
  publisher={Optica Publishing Group}
}

@article{pupeza2007highly,
  title={Highly accurate optical material parameter determination with THz time-domain spectroscopy},
  author={Pupeza, Ioachim and Wilk, Rafal and Koch, Martin},
  journal={Optics express},
  volume={15},
  number={7},
  pages={4335--4350},
  year={2007},
  publisher={Optica Publishing Group}
}

@article{wilk2008highly,
  title={Highly accurate THz time-domain spectroscopy of multilayer structures},
  author={Wilk, Rafal and Pupeza, Ioachin and Cernat, Radu and Koch, Martin},
  journal={IEEE Journal of Selected Topics in Quantum Electronics},
  volume={14},
  number={2},
  pages={392--398},
  year={2008},
  publisher={IEEE}
}

@inproceedings{yang2024theoretical,
  title={Theoretical Models for Performance Analysis of Spintronic THz Emitters},
  author={Yang, Yingshu and Dal Forno, Stefano and Battiato, Marco},
  booktitle={Photonics},
  volume={11},
  number={8},
  pages={730},
  year={2024},
  organization={MDPI}
}

@article{yang2024perturbative,
  title={Perturbative transfer matrix method for optical-pump terahertz-probe spectroscopy of ultrafast dynamics in spintronic terahertz emitters},
  author={Yang, Yingshu and Dal Forno, Stefano and Battiato, Marco},
  journal={Physical Review B},
  volume={109},
  number={2},
  pages={024425},
  year={2024},
  publisher={APS}
}

@article{yang2023modeling,
  title={Modeling spintronic terahertz emitters as a function of spin generation and diffusion geometry},
  author={Yang, Yingshu and Dal Forno, Stefano and Battiato, Marco},
  journal={Physical Review B},
  volume={107},
  number={14},
  pages={144407},
  year={2023},
  publisher={APS}
}

@article{yang2021removal,
  title={Removal of spectral distortion due to echo for ultrashort THz pulses propagating through multilayer structures with thick substrate},
  author={Yang, Yingshu and Dal Forno, Stefano and Battiato, Marco},
  journal={Journal of Infrared, Millimeter, and Terahertz Waves},
  pages={1--11},
  year={2021},
  publisher={Springer}
}

@article{yang2021transfer,
  title={Transfer-matrix description of heterostructured spintronics terahertz emitters},
  author={Yang, Yingshu and Dal Forno, Stefano and Battiato, Marco},
  journal={Physical Review B},
  volume={104},
  number={15},
  pages={155437},
  year={2021},
  publisher={APS}
}

@article{liu2022spintronic,
  title={Spintronic terahertz emitters in silicon-based heterostructures},
  author={Liu, Jiayun and Lee, Kyusup and Yang, Yingshu and Li, Ziqi and Sharma, Raghav and Xi, Lifei and Salim, Teddy and Boothroyd, Chris and Lam, Yeng Ming and Yang, Hyunsoo and others},
  journal={Physical Review Applied},
  volume={18},
  number={3},
  pages={034056},
  year={2022},
  publisher={APS}
}

@article{agarwal2023secondary,
  title={Secondary spin current driven efficient THz spintronic emitters},
  author={Agarwal, Piyush and Yang, Yingshu and Medwal, Rohit and Asada, Hironori and Fukuma, Yasuhiro and Battiato, Marco and Singh, Ranjan},
  journal={Advanced Optical Materials},
  volume={11},
  number={23},
  pages={2301027},
  year={2023},
  publisher={Wiley Online Library}
}

@article{agarwal2022terahertz,
  title={Terahertz spintronic magnetometer (TSM)},
  author={Agarwal, Piyush and Yang, Yingshu and Lourembam, James and Medwal, Rohit and Battiato, Marco and Singh, Ranjan},
  journal={Applied Physics Letters},
  volume={120},
  number={16},
  year={2022},
  publisher={AIP Publishing}
}

@article{kampfrath2013terahertz,
  title={Terahertz spin current pulses controlled by magnetic heterostructures},
  author={Kampfrath, Tobias and Battiato, Marco and Maldonado, Pablo and Eilers, Gerrit and N{\"o}tzold, J and M{\"a}hrlein, Sebastian and Zbarsky, Vladyslav and Freimuth, Frank and Mokrousov, Yuriy and Bl{\"u}gel, Stefan and others},
  journal={Nature nanotechnology},
  volume={8},
  number={4},
  pages={256--260},
  year={2013},
  publisher={Nature Publishing Group}
}

@article{seifert2016efficient,
  title={Efficient metallic spintronic emitters of ultrabroadband terahertz radiation},
  author={Seifert, Tom and Jaiswal, S and Martens, U and Hannegan, J and Braun, Lukas and Maldonado, Pablo and Freimuth, F and Kronenberg, A and Henrizi, J and Radu, I and others},
  journal={Nature photonics},
  volume={10},
  number={7},
  pages={483--488},
  year={2016},
  publisher={Nature Publishing Group UK London}
}

@article{qiu2018layer,
  title={Layer thickness dependence of the terahertz emission based on spin current in ferromagnetic heterostructures},
  author={Qiu, HS and Kato, K and Hirota, K and Sarukura, N and Yoshimura, M and Nakajima, M},
  journal={Optics Express},
  volume={26},
  number={12},
  pages={15247--15254},
  year={2018},
  publisher={Optica Publishing Group}
}

@article{torosyan2018optimized,
  title={Optimized spintronic terahertz emitters based on epitaxial grown Fe/Pt layer structures},
  author={Torosyan, Garik and Keller, Sascha and Scheuer, Laura and Beigang, Ren{\'e} and Papaioannou, Evangelos Th},
  journal={Scientific reports},
  volume={8},
  number={1},
  pages={1311},
  year={2018},
  publisher={Nature Publishing Group UK London}
}

@article{zhou2018broadband,
  title={Broadband terahertz generation via the interface inverse Rashba-Edelstein effect},
  author={Zhou, C and Liu, YP and Wang, Z and Ma, SJ and Jia, M Wu and Wu, RQ and Zhou, L and Zhang, W and Liu, MK and Wu, YZ and others},
  journal={Physical review letters},
  volume={121},
  number={8},
  pages={086801},
  year={2018},
  publisher={APS}
}

@article{ordal1985optical,
  title={Optical properties of fourteen metals in the infrared and far infrared: Al, Co, Cu, Au, Fe, Pb, Mo, Ni, Pd, Pt, Ag, Ti, V, and W.},
  author={Ordal, Mark A and Bell, Robert J and Alexander, Ralph W and Long, Larry L and Querry, Marvin R},
  journal={Applied optics},
  volume={24},
  number={24},
  pages={4493--4499},
  year={1985},
  publisher={Optica Publishing Group}
}

@article{bull2021spintronic,
  title={Spintronic terahertz emitters: Status and prospects from a materials perspective},
  author={Bull, Charlotte and Hewett, Simmone M and Ji, Ruidong and Lin, Cheng-Han and Thomson, Thomas and Graham, Darren M and Nutter, Paul W},
  journal={Apl Materials},
  volume={9},
  number={9},
  year={2021},
  publisher={AIP Publishing}
}

@inproceedings{franta2016optical,
  title={Optical characterization of SiO2 thin films using universal dispersion model over wide spectral range},
  author={Franta, Daniel and Ne{\v{c}}as, David and Ohl{\'\i}dal, Ivan and Giglia, Angelo},
  booktitle={Optical Micro-and Nanometrology VI},
  volume={9890},
  pages={253--267},
  year={2016},
  organization={SPIE}
}

@article{koenig2013wireless,
  title={Wireless sub-THz communication system with high data rate},
  author={Koenig, Swen and Lopez-Diaz, Daniel and Antes, Jochen and Boes, Florian and Henneberger, Ralf and Leuther, Arnulf and Tessmann, Axel and Schmogrow, Ren{\'e} and Hillerkuss, David and Palmer, Robert and others},
  journal={Nature photonics},
  volume={7},
  number={12},
  pages={977--981},
  year={2013},
  publisher={Nature Publishing Group UK London}
}

@article{chen2020ghost,
  title={Ghost spintronic THz-emitter-array microscope},
  author={Chen, Si-Chao and Feng, Zheng and Li, Jiang and Tan, Wei and Du, Liang-Hui and Cai, Jianwang and Ma, Yuncan and He, Kang and Ding, Haifeng and Zhai, Zhao-Hui and others},
  journal={Light: Science \& Applications},
  volume={9},
  number={1},
  pages={99},
  year={2020},
  publisher={Nature Publishing Group UK London}
}

@article{luo2020nanoscale,
  title={Nanoscale terahertz STM imaging of a metal surface},
  author={Luo, Yang and Jelic, Vedran and Chen, Gong and Nguyen, Peter H and Liu, Yu-Jui Ray and Calzada, Jesus AM and Mildenberger, Daniel J and Hegmann, Frank A},
  journal={Physical Review B},
  volume={102},
  number={20},
  pages={205417},
  year={2020},
  publisher={APS}
}

@misc{beard2002terahertz,
  title={Terahertz spectroscopy},
  author={Beard, Matthew C and Turner, Gordon M and Schmuttenmaer, Charles A},
  journal={The Journal of Physical Chemistry B},
  volume={106},
  number={29},
  pages={7146--7159},
  year={2002},
  publisher={ACS Publications}
}

@article{cheng2019far,
  title={Far out-of-equilibrium spin populations trigger giant spin injection into atomically thin MoS2},
  author={Cheng, Liang and Wang, Xinbo and Yang, Weifeng and Chai, Jianwei and Yang, Ming and Chen, Mengji and Wu, Yang and Chen, Xiaoxuan and Chi, Dongzhi and Goh, Kuan Eng Johnson and others},
  journal={Nature Physics},
  volume={15},
  number={4},
  pages={347--351},
  year={2019},
  publisher={Nature Publishing Group UK London}
}

@article{dorney2001material,
  title={Material parameter estimation with terahertz time-domain spectroscopy},
  author={Dorney, Timothy D and Baraniuk, Richard G and Mittleman, Daniel M},
  journal={Journal of the Optical Society of America A},
  volume={18},
  number={7},
  pages={1562--1571},
  year={2001},
  publisher={Optical Society of America}
}

@article{la2017phonon,
  title={Phonon features in terahertz photoconductivity spectra due to data analysis artifact: A case study on organometallic halide perovskites},
  author={La-o-Vorakiat, Chan and Cheng, Liang and Salim, Teddy and Marcus, Rudolph A and Michel-Beyerle, Maria-Elisabeth and Lam, Yeng Ming and Chia, Elbert EM},
  journal={Applied Physics Letters},
  volume={110},
  number={12},
  year={2017},
  publisher={AIP Publishing}
}

@misc{suppinfo,
  title={Supplementary information},
  author={Yingshu Yang},
  note={See Supplemental Material at [url] for algorithmic details, input specifications, and additional fitting results.}
}

@article{beermann2025terahertz,
  title={Terahertz time-domain spectroscopy for simultaneous measurement of optical constants, and material thickness with deep-subwavelength precision},
  author={Beermann, Nicolas S and Gebauer, Andreas and Fabretti, Savio and Zhang, Wentao and Hiraoka, Tomoki and Achtstein, Alexander W and Hafez, Hassan A and Turchinovich, Dmitry},
  journal={Optics Express},
  volume={33},
  number={4},
  pages={8650--8660},
  year={2025},
  publisher={Optica Publishing Group}
}

@article{li2025electric,
  title={Electric-Field Tunable THz Emission via Quantum Geometry in Dirac Semimetal},
  author={Li, Ziqi and Yang, Dongsheng and Wang, Fei and Yang, Yingshu and Guo, Yuanyuan and Bao, Di and Yin, Tingting and Tang, Chi Sin and Salim, Teddy and Xi, Lifei and others},
  journal={Nano Letters},
  volume={25},
  number={22},
  pages={9006--9014},
  year={2025},
  publisher={ACS Publications}
}

@article{ordal1983optical,
  title={Optical properties of the metals al, co, cu, au, fe, pb, ni, pd, pt, ag, ti, and w in the infrared and far infrared},
  author={Ordal, Mark A and Long, LL and Bell, Robert John and Bell, SE and Bell, RR and Alexander, Ralph William and Ward, CA},
  journal={Applied optics},
  volume={22},
  number={7},
  pages={1099--1119},
  year={1983},
  publisher={OSA}
}

@article{blaber2009search,
  title={Search for the ideal plasmonic nanoshell: the effects of surface scattering and alternatives to gold and silver},
  author={Blaber, Martin G and Arnold, Matthew D and Ford, Michael J},
  journal={The Journal of Physical Chemistry C},
  volume={113},
  number={8},
  pages={3041--3045},
  year={2009},
  publisher={ACS Publications}
}

@article{zeman1987accurate,
  title={An accurate electromagnetic theory study of surface enhancement factors for silver, gold, copper, lithium, sodium, aluminum, gallium, indium, zinc, and cadmium},
  author={Zeman, Ellen J and Schatz, George C},
  journal={Journal of Physical Chemistry},
  volume={91},
  number={3},
  pages={634--643},
  year={1987},
  publisher={ACS Publications}
}

@article{hooper2002dispersion,
  title={Dispersion of surface plasmon polaritons on short-pitch metal gratings},
  author={Hooper, Ian R and Sambles, J Roy},
  journal={Physical Review B},
  volume={65},
  number={16},
  pages={165432},
  year={2002},
  publisher={APS}
}

@article{grady2004influence,
  title={Influence of dielectric function properties on the optical response of plasmon resonant metallic nanoparticles},
  author={Grady, Nathaniel K and Halas, Naomi J and Nordlander, Peter},
  journal={Chemical Physics Letters},
  volume={399},
  number={1-3},
  pages={167--171},
  year={2004},
  publisher={Elsevier}
}

@article{seifert2018terahertz,
  title={Terahertz spectroscopy for all-optical spintronic characterization of the spin-Hall-effect metals Pt, W and Cu80Ir20},
  author={Seifert, Tom S and Tran, Ngoc Minh and Gueckstock, Oliver and Rouzegar, Seyed Mohammedreza and Nadvornik, Lukas and Jaiswal, Samridh and Jakob, Gerhard and Temnov, Vasily V and M{\"u}nzenberg, Markus and Wolf, Martin and others},
  journal={Journal of Physics D: Applied Physics},
  volume={51},
  number={36},
  pages={364003},
  year={2018},
  publisher={IOP Publishing}
}

@article{rakic1998optical,
  title={Optical properties of metallic films for vertical-cavity optoelectronic devices},
  author={Raki{\'c}, Aleksandar D and Djuri{\v{s}}i{\'c}, Aleksandra B and Elazar, Jovan M and Majewski, Marian L},
  journal={Applied optics},
  volume={37},
  number={22},
  pages={5271--5283},
  year={1998},
  publisher={OSA}
}

\end{document}